# Automatic Electrodes Detection during simultaneous EEG/fMRI acquisition

Mathis Fleury[1], Pierre Maurel[1], Marsel Mano[1], Elise Bannier[2], and Christian Barillot[1]

[1]Univ Rennes, INRIA, CNRS, INSERM, VISAGES-ERL U 1228, F-35000 Rennes FRANCE, Rennes, France, [2]Univ Rennes, INRIA, CNRS, INSERM, VISAGES-ERL U 1228, F-35000 Rennes FRANCE, University Hospital of Rennes Department Radiology, Rennes, France

## Synopsis

Simultaneous EEG/fMRI acquisition allows to measure brain activity at high spatial-temporal resolution. The localisation of EEG sources depends on several parameters including the position of the electrodes on the scalp. The position of the MR electrodes during its acquisitions is obtained with the use of the UTE sequence allowing their visualisation. The retrieval of the electrodes consists in obtaining the volume where the electrodes are located by applying a sphere detection algorithm. We detect around 90% of electrodes for each subject, and our UTE-based electrode detection showed an average position error of 3.7mm for all subjects.

## Purpose

The coupling of EEG/fMRI allows to measure brain activity at high spatial-temporal resolution. The localisation of EEG sources depends on several parameters including the position of the electrodes on the scalp. An accurate knowledge about these informations is important for source reconstruction. Currently, when acquiring EEG/fMRI together, the position of the electrodes is generally estimated according to fiducial points by using a template. In the context of simultaneous EEG/fMRI acquisition, a natural idea is to use MR images to localise EEG electrodes. However MR compatible electrodes are built to be almost invisible on MRI Images. Taking advantage of a recently proposed Ultra-Short Echo Time (UTE) sequence [2], we propose a fully automatic method to detect those electrodes in MR images.

## Materials & methods

The UTE sequence was acquired in five healthy volunteers and all underwent a simultaneous EEG/fMRI examination. A 64-channels MR compatible cap (Brainproduct, Gilching, Germany) was used for all the subjects. The retrieval of the electrodes consisted in two part, in the one hand we provided a mask that consists in obtaining the volume where the electrodes are located. In the second hand we performed the electrode detection inside the volume of interest (VOI).

First, the anatomical T1 image is segmented using FSL [3], and a mask of the scalp is computed from the segmentation. This mask is then dilated to isolate the volume where the electrodes are located. Figure 2 shows the result of this first step.

The second step consists in obtaining a list of potential candidates for electrodes position. This list is given by a spherical Hough algorithm that will detect spherical shapes in our VOI. Detection of spherical shapes are presented in figure 3.a.

Third, we will now filter these potential electrodes to keep only the real electrodes by applying a criterion based on their theoretical positions. To do this, we use a spherical 64-template EEG net and the Iterative Closest Point algorithm (ICP) to register the template to the detected points. For each of the 64 electrodes of the registered template, the closest point detected by the spherical Hough is selected. Unselected points are discarded. Once the filtering step has been performed the number of electrodes will be equal to 64. Detected electrodes for one subject are shown in figure 3.b.

Finally the points that are far from the registered template are discarded and replaced by the local maxima of the UTE image around the theoretical position given by the registered template. The 64 obtained detection are then labelled using the template.

## Results

The performance of our method was evaluated by comparing our detections to a ground truth (GT). Following [1], this GT was manually obtained by picking up the position of each 64 electrodes for each subject on the pancake view, which is roughly a 2D projection of the scalp.

We also compared our method to a more classical semi automatic one [5]: five fiducial points were obtained manually and the spherical template was adjusted to these points.

The figures 4 and 5 illustrate that we are able to detect almost 90% of electrodes for each subject. Our UTE-based electrode detection showed an average position error of 3.7mm for all subjects. Paired t-test was computed between the two groups with $p < 0.0001$.

## Discussion

We have shown that our method offers constant and precise results for each subject. Moreover, this method of detection based-MR provides the position of the electrodes directly into the MR-space.

Today, the time for a UTE sequence is almost five minutes. Lowering the sample of this sequence makes the acquisition time can going down to 1 min 30. We expect to evaluate the impact of this down-sampling in the near future.

### Conclusion

We present a method to detect and label automatically EEG electrodes (64-channels) during an EEG/fMRI acquisition. We have used a UTE sequence to obtain the electrode position on a MR-volume. This method does not require additional cost and can be used by just adding a UTE sequence to the MR protocol.

For future research, this technique will be used to extract the position of the electrodes in real time. Indeed, this technique is interesting for applications requiring immediate knowledge of the position of the electrodes. We believe this method will be useful in the improvement of the fusion of EEG-fMRI data-sets.

### Acknowledgements

This work has received a French government support granted to the CominLabs excellence laboratory and managed by the National Research Agency in the "Investing for the Future" program under reference ANR-10-LABX-07-01. It was also financed by Brittany region under HEMISFER project, MRI data acquisition was supported by the Neurinfo MRI research facility from the University of Rennes I. Neurinfo is granted by the European Union (FEDER),the French State, the Brittany Council, Rennes.

### Figures

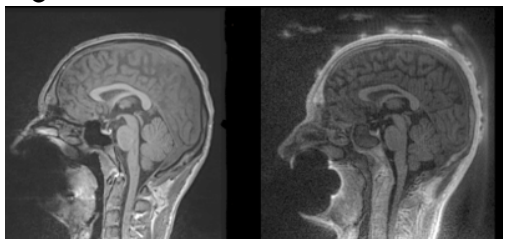

Figure 1: Visualisation of the electrodes around the scalp, the T1 image (on the right) show that the MR compatible electrodes does not appear however they emerge on the UTE sequence.

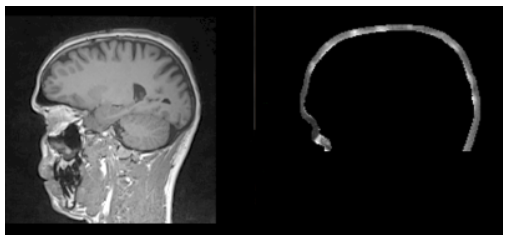

Figure 2: Extraction of the VOI on the T1 image (right) and UTE image masked with the VOI.

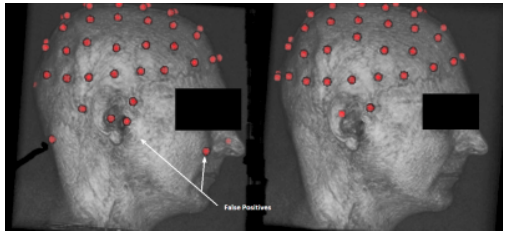

Figure 3: Example of electrode detection before (on the right) appliance of the filtering and after filtering.

| Subjects | FN | FP | detection accuracy (%) |
|---|---|---|---|
| S1 | 9 | 9 | 86.2 |
| S2 | 4 | 4 | 93.8 |
| S3 | 2 | 2 | 96.9 |
| S4 | 7 | 7 | 89.2 |
| S5 | 8 | 8 | 87.7 |

Figure 4: Number of false positive (FP) and false negative (FN) for every subject. The detection accuracy is the percentage of electrodes that have been detected. We consider that an electrode is detected when the position error (PE) is bellow 1cm (Kavanagk al. 1978 [4]).

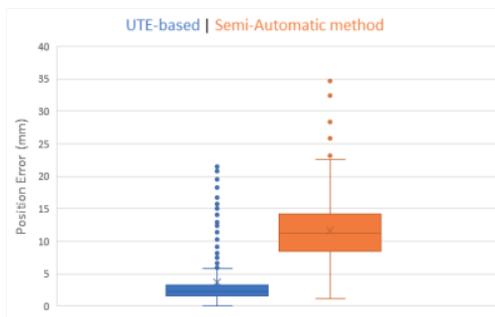

Figure 5: Position error for UTE-based detection and semi-automatic detection for all subjects.